\documentclass[12pt]{article}
\usepackage{epsfig,amsmath,amsfonts}
\usepackage[figuresright]{rotating}
\newcommand{\bm}[1]{\mbox{\boldmath $#1$}}
\begin{document}
\title{The Richardson's Law in Large-Eddy Simulations 
of Boundary Layer flows}
\author{G. Gioia$^1$, G. Lacorata$^{1}$, E.P. Marques Filho$^{2}$,\\
A. Mazzino$^{1,3}$ and U.Rizza$^{1}$\\
\small{$^1$ ISAC-CNR, Sezione di Lecce, I--73100, Lecce, Italy}\\
\small{$^2$ Institute of Astronomy, Geophysics and Atmospheric
Sciences,}\\
\small{University of Sao Paulo, 05508--900, Sao Paulo, Brasil}\\
\small{$3$ Dipartimento di Fisica, Universit\`a di Genova, I--16146,
Genova, Italy}}
\date{\today}
\maketitle
\begin{abstract}
Relative dispersion in a neutrally stratified planetary boundary layer
(PBL) is investigated by means of Large-Eddy Simulations (LES).
Despite the small extension of the inertial range of scales
in the simulated PBL,
our Lagrangian statistics  turns out to be compatible with the
Richardson $t^3$ law for the average of square particle separation.
This emerges from the application of nonstandard methods of analysis
through which a precise measure of the Richardson constant
was also possible. Its values is estimated as $C_2\sim 0.5$ in 
close agreement with recent experiments and three-dimensional direct
numerical simulations.
\end{abstract}
\section{Introduction}  
One of the most striking features of a turbulent planetary
boundary layer (PBL) is the presence of a wide range of active length
scales. They range from the smallest dynamically active scales
of the order of millimeters (the so-called Kolmogorov scale),
below which diffusive effects are dominant, to the largest 
scales of the order of  ten kilometers.  Such a large range of
excited scales are essentially a continuum and the distribution of energy 
scale-by-scale is controlled by the famous Kolmogorov's 1941 prediction 
(see Frisch, 1995 for a modern presentation).\\
One of the most powerful concepts which highlighted the dynamical
role of the active scales in the atmosphere 
was due to Richardson (1926). He introduced in
his pioneering work the concept of  turbulent relative dispersion
(see Sawford, 2001 for a recent review)
with the aim of investigating the large variations of atmospheric turbulent 
diffusion when observed at different spatial scales. \\
In his work, Richardson proposed a diffusion equation 
for the probability density function, $p({\bm r},t)$,
 of pair separation. Assuming isotropy such an equation can be cast 
into the form
\begin{equation}
\frac{\partial p({\bm r},t)}{\partial t} = \frac{1}{r^2}
\frac{\partial  }{\partial r}\left [ r^2 D(r)
\frac{\partial p(\bm{r},t)}{\partial r} \right ]
\label{diffusion}
\end{equation}
where the scale-dependent eddy-diffusivity $D(r)$  
 accounts for the
enormous increase in observed values of the turbulent diffusivity 
in the atmosphere.\\ The famous scaling law $D(r) \propto r^{4/3}$ 
was obtained by Richardson (1926) from experimental data. 
 From the expression of $D(r)$ as function of $r$ 
and exploiting Eq.~(\ref{diffusion})
the well known non-Gaussian distribution
\begin{equation}
p({\bm r},t) \propto t^{-9/2}
\exp\left(- C r^{2/3}/t \right)
\label{eq:2.2}
\end{equation}
is easily obtained.\\
This equation implies that 
the mean square particle separation grows as
\begin{equation}
R^2(t) \equiv \langle r^2(t) \rangle = C_2 \epsilon t^{3}
\label{eq:2.3}
\end{equation}
which is the celebrated Richardson's  
 ``$t^3$'' law for the pair dispersion. Here $C_2$ is  
the so-called Richardson constant and  $\epsilon$ is the mean 
energy dissipation. \\
Despite the fact that the Richardson's law has been proposed 
since a long time,  there is still a large uncertainty on the value
of $C_2$. Some authors have found $C_2$ ranging from  $\sim 10^{-2}$
to $\sim 10^{-1}$  in kinematic simulations 
(see, for example, Elliot and Majda, 1996; Fung and Vassilicos, 1998), 
although 
for kinematic models an energy flux $\epsilon$ can hardly be defined.   
On the other hand, a value $C_2 \sim O(1)$ (and even larger) 
 follows from closure predictions (Monin and Yaglom, 1975). 
More recently, both an experimental investigation (Ott and  Mann, 2000)
and accurate three-dimensional direct numerical simulations (DNS) 
(Boffetta and Sokolov, 2002) give a strong support for the value 
$C_2 \sim 0.5$.\\
The main limitation of the state-of-the-art three-dimensional DNS 
is that the achieved Reynolds numbers are still far from those
characterizing the so-called fully developed turbulence regime, that is
the realm of  the Richardson's (1926) theory. 
Moreover, initial and boundary
conditions assumed in the most advanced DNS are, however, quite idealized 
and do not match those characterizing a turbulent PBL, 
the main concern of the present paper.\\
For all these reasons we have decided to focus our attention on
Large-Eddy Simulations (LES) of a neutrally stratified PBL
and address the issue related to the determination of
the Richardson constant $C_2$.
The main advantage of this strategy is that it permits
to achieve very high Reynolds numbers and, at the same time, 
it properly reproduces the dynamical features observed in the PBL.\\
It is worth anticipating that the naive approach which should 
lead to the determination of $C_2$ by looking at  
 the behavior of $R^2(t)$ {\it versus} the time $t$ 
is extremely sensitive to the initial pair separations
and thus gives estimations of the Richardson's constant
which appear quite questionable (see Fig.~3).
This is simply due to the fact that, in realistic situations like
the one we consider, the inertial range of scales is quite
narrow and, consequently, there is no room for a genuine $t^3$ regime
to appear (see Boffetta et al., 2000 for 
general considerations on this important point). \\
This fact motivated  us to apply a recently established  
`nonstandard' analysis technique (the so-called FSLE approach, 
Boffetta et al., 2000)
to isolate a clear Richardson regime and thus to provide a reliable
and systematic (that is independent from initial pair separations)
measure for $C_2$. This is the main aim of our paper.

\section{The LES strategy}
%code and the numerical experiments}
In a  LES strategy the large scale motion 
(that is motion associated to the largest turbulent eddies) is explicitly solved 
while the smallest scales (typically in the inertial range of scales)
are described in a statistical consistent way
(that is parameterized in terms of the resolved, large scale, 
velocity and temperature fields).
This is done by filtering the governing equations for velocity and
potential temperature
by means of a filter operator. Applied, for example, to the
$i$th-component of the velocity field, $u_i$,
($u_1=u$, $u_2=v$, $u_3=w$), the filter
is defined by the convolution:
\begin{equation}
\overline{u}_i({\bm x})=\int u_i({\bm x}')G({\bm x}-{\bm x}')d{\bm x}'
\end{equation}
where $\overline{u}_i$ is the filtered field and $G({\bm x})$ is a
three-dimensional filter function. The field component $u_i$ can be
thus decomposed as 
\begin{eqnarray}
\label{decom}
u_i=\overline{u}_i+{u}_i'' 
\end{eqnarray}
and similarly for the temperature field.
In our model, the equation for the latter field is 
coupled to the Navier--Stokes equation via the Boussinesq term.\\ 
Applying the filter operator both to the Navier--Stokes equation 
and to the equation for the potential temperature, and exploiting 
the decomposition (\ref{decom}) (and the analogous for the temperature field)
in the advection terms one 
obtains the corresponding filtered equations:
\begin{eqnarray}
\frac{\partial \overline{u}_i}{\partial t} &=& -\frac{\partial
\overline{\overline{u}_i\overline{u}_j}}{\partial x_j} -
\frac{\partial\tau_{ij}^{(u)}}{\partial x_j}
-\frac{1}{\rho}\frac{\partial \overline{p}}{\partial x_i } + g_i 
\frac{\overline{\theta}}{\theta_0}\delta_{i3}
-f\epsilon_{ij3}\overline{u}_j+ \nu\nabla^2\overline{u}_i
\label{filt1}\\
\frac{\partial \overline{u}_i}{\partial x_i} &=& 0 \label{filt2} \\
\frac{\partial \overline{\theta}}{\partial t} &=&-\frac{\partial
\overline{\overline{u_j}\overline{\theta}}}{\partial x_j}-
\frac{\partial\tau_{j}^{(\theta)}}{\partial x_j}+\kappa\nabla^2\overline{\theta}
\label{filt3}
\end{eqnarray}
where
$\rho$ is the air density, $p$ is the pressure, $f$ is the Coriolis
parameter, $\nu$ is the molecular viscosity, $\kappa$ is the thermal
molecular diffusivity, $ g_i \frac {\theta}{\theta_0}\delta_{i3}$ is
the buoyancy term and $\theta_0$ is a reference temperature profile.
The quantities to be parametrized in terms of large scale fields are
\begin{equation}
\tau_{ij}^{(u)}=\overline{\overline{u}_i u''_j}+\overline{u''_i\overline{u}_j
}+\overline{u''_i u''_j};\qquad
 \tau_{j}^{(\theta)}=\overline{\overline{\theta}u''_j}+
\overline{\theta''\overline{u}_j}+\overline{\theta'' u''_j} ,
\end{equation}
that represent the subgrid scale (SGS) fluxes of momentum and heat, 
respectively.\\
\begin{table} 
\caption[]{The relevant parameters characterizing the simulated
PBL. In this table, $L_x$, $L_y$
and $L_z$ are the domain extension along the directions $x$, $y$ and
$z$, respectively;  $Q_*$ is the heat flux from the bottom boundary; $U_g$ is
the geostrophic wind; $z_i$ is  the mixed layer depth, $u_*$ is the 
friction velocity and $\tau_*\equiv z_i/u_*$ is the
turnover time;} 
\begin{tabular}{lrll}                                       
\hline
\multicolumn{2}{l}{\it parameter} & {\it value}\\ 
\hline  
$L_{x}$, $L_{y}$           & [km]  &  2               \\
$L_{z}$             & [km]     & 1                 \\
$Q_*$             & [m K s$^{-1}$]        & 0             \\
$U_g$             & [m s$^{-1}$]        &  15                \\
$z_i$           & [m]     & 461                    \\
$u_*$           & [ms$^{-1}$]     & 0.7            \\
$\tau_*$           & [s]  &  674                \\
\hline
\end{tabular}
\end{table}
In our model:
\begin{equation}
\tau^{(u)}_{i j}=-2 K_{M}\left ( \partial_{i} \overline{u}_{j} +  
\partial_{j} \overline{u}_{i})
\right )
\end{equation}
\begin{equation}
\tau^{(\theta)}_{i}=-K_{H}\partial_{i}\overline{\theta}
\end{equation}
$K_{M}$ and $K_{H}$ being the SGS eddy coefficients for momentum and
heat, respectively.\\
The above two eddy coefficients are related to the velocity scale
$\overline{e'}^{1/2}$, $\overline{e'}$ being the SGS turbulence energy the 
equation of which is solved in our LES model (Moeng, 1984), 
and to the length scale
$l\equiv (\Delta x\Delta y\Delta z )^{1/3}$ (valid for neutrally stratified
cases) $\Delta x$, $\Delta y$, and $ \Delta z $ being the grid mesh
spacing in $x$, $y$ and $z$. Namely:
\begin{equation}
K_{M}=0.1\; l\; \overline{e'}^{1/2}
\end{equation}
\begin{equation}
K_{H}=3 K_{M}.  
\end{equation}
Details on the  LES model we used in our study can be found
in Moeng, 1984 
and in Sullivan et al., 1994. Such a model 
has been widely used and tested to investigate basic research
problems in the framework of boundary layer flows
(see, for example, Antonelli et al., 2003 and Moeng and Sullivan, 1994 
among the others). 
\section{The simulated PBL}
In order to obtain a stationary PBL we advanced in time
our LES code for around six large-eddy turnover times, $\tau_*$,
with a spatial resolution of $128^3$ grid points. This time
will be the starting point for the successive Lagrangian analysis
(see next section).\\
The relevant parameters characterizing our simulated PBL
are listed in Table 1 at $t=6\;\tau_*$. At the same instant,
%For a sensitivity test,
%simulations with the resolution  $96^3$ have been also performed. \\
we show in Fig.~1 the horizontally averaged vertical profile of the velocity
components $u$, $v$. The average of the
vertical component is not shown, the latter being 
very close to zero.
We can observe the presence of a rather well 
mixed region which extends from $z\sim 0.2\;z_i$  to $z\sim  z_i$.
\begin{figure}
\begin{center}
\includegraphics[width=12cm,height=7cm]{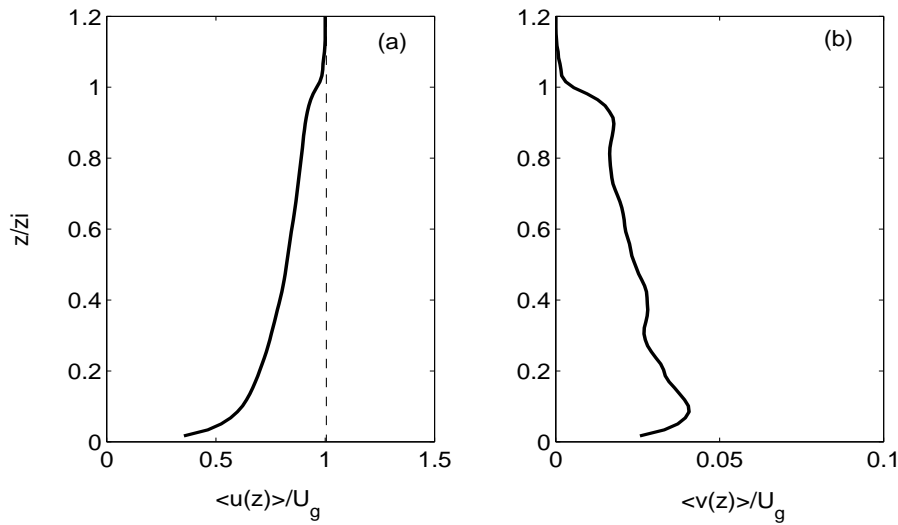}
\caption{The horizontally averaged velocity profiles.
(a): stream-wise $u$-component, (b) span-wise $v$-component.} 
\end{center}
\end{figure}
The energy spectra for the three velocity components are reported
in Fig.~2. Dashed lines are relative to the Kolmogorov (K41)
prediction $E(k)\propto k^{-5/3}$.  Although 
the inertial range of scale appears quite narrow, data  
are compatible with the K41 prediction.
\begin{figure}
\begin{center}
\includegraphics[width=12cm,height=5cm]{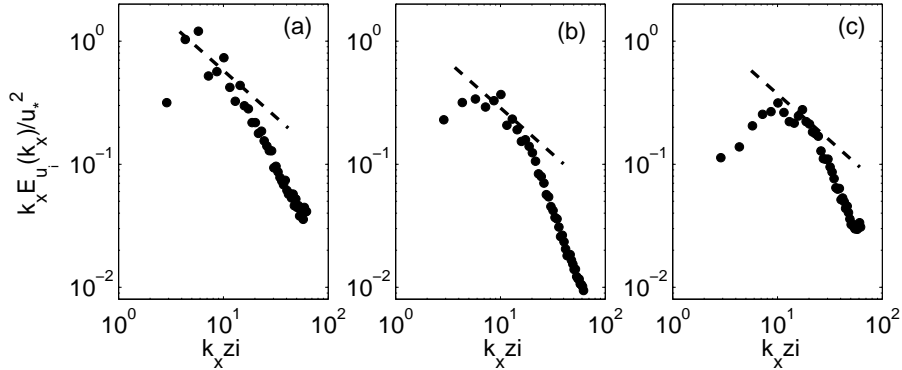}
\caption{Energy spectra for the three components of the velocity field.
(a): stream-wise, (b) span-wise, (c) vertical.  The dashed lines
correspond to the K41 prediction.}
\end{center}
\end{figure}

\section{Lagrangian simulations}
In order to investigate the statistics of pair dispersion,
from the time $t=6\;\tau_*$ (corresponding to the PBL stationary state)
we integrated, in parallel to the LES,
the equation for the passive tracer trajectories
defined by the equation
\begin{equation}
\dot{\bm x}(t) = {\bm u}({\bm x}(t),t) .
\label{traj}
\end{equation}
We performed a single long run where the evolution of 
20000 pairs has been followed starting from two different
initial separations: $R(0)=\Delta x$ and $R(0)=2 \Delta x$, $\Delta x$
being the
grid mesh spacing  whose value is 15.6 $m$.
Trajectories have been integrated for a  time 
of the order of 5000 $s$ with a time step of around 1 $s$, 
the same used to advance in time the LES.\\
At the initial time, pairs are  uniformly distributed 
on a horizontal plane placed at the elevation $z_i/2$.
Reflection has been assumed both at the 
capping inversion (at the elevation $z_i$)
and at the bottom boundary.\\
For testing purposes, a second run (again started from  $t=6\;\tau_*$) 
with a smaller number of pairs
(5000) has been performed. No significant differences 
in the Lagrangian statistics have been however observed. The same
conclusion has been obtained 
for a second test where the LES spatial resolution has been lowered
to $96^3$ grid points. For a comparison see Figs.~3 and 4.\\
The velocity field necessary to integrate (\ref{traj})
has been  obtained by a bilinear interpolation
from the eight nearest grid points on which the velocity field produced
by the  LES  is defined.\\
In this preliminary investigation, we did not use any sub-grid
model describing the Lagrangian contribution arising from
the motion on scales smaller than the grid mesh spacing.\\ 

\subsection{Pair dispersion statistics}
In Fig.~3 we show the second moment of relative dispersion
$R^2(t)$ for the two initial separations. Heavy dashed line
represents the expected Richardson's law, which is however
not compatible with our data for the largest initial separation $2\Delta x$. 
We can also notice how the $R^2(t)$ curve becomes flatter for larger
separations.  The same 
dependence has been observed by Boffetta and Celani (2000) for pair
dispersion in two-dimensional turbulence. \\
\begin{figure}
\begin{center}
\includegraphics[width=12cm,height=8cm]{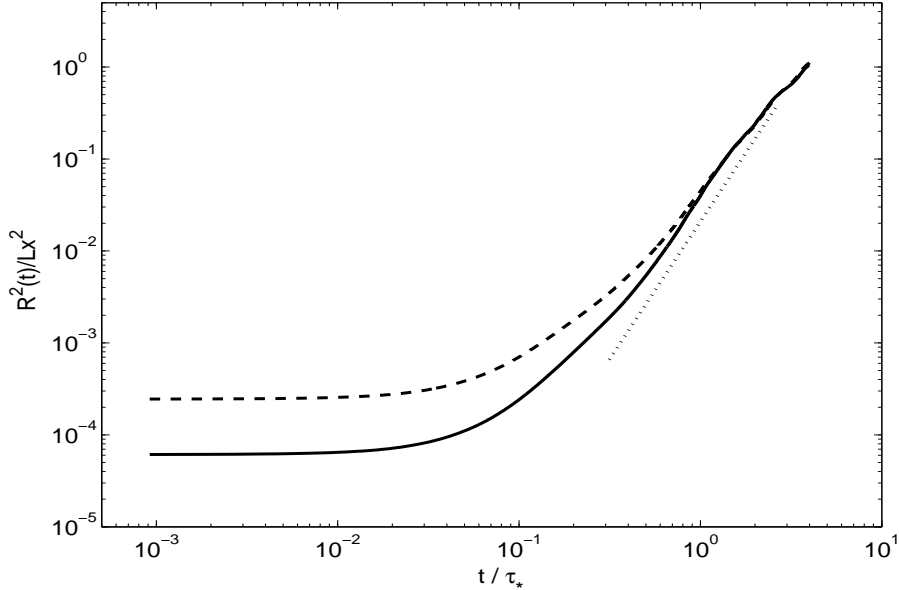}
\caption{The behavior of the (dimensionless) mean square relative dispersion
 {\it vs}  the (dimensionless) time. Full line: the initial separation
is $\Delta x$; Dashed-line: 
the initial separation is $2 \Delta x$. Dotted line
is relative to the $t^3$ Richardson's law.}
\end{center}
\end{figure}
The fact that our data do not fit
 the Richardson law, for generic initial pair separations,
is simply explained as a consequence of finite size effects
(in space and in time) of our system. Indeed, 
it is clear that, unless $t$ is large enough that all particle pairs have
``forgotten'' their initial conditions, the average will be biased.
This is why we observe a consistent flattening of $R^2(t)$
at small times. Such regime is a crossover from initial
conditions to the Richardson regime.  From Fig.~3 we can see
that the extension of such crossover increases as the initial
separation increases.\\
Unfortunately,  we cannot augment the time $t$ too much
because of the reduced extension of our inertial range (see Fig.~2).
To overcome this problem, and thus to allow a systematic estimation
of the Richardson constant which does not depend on the choice of the initial
pair separation,   we use an alternative approach
based on statistics at {\it fixed scale} (Boffetta et al., 2000). 
This is the subject of the next subsection.

\subsection{Fixed-scale statistics}

The  characterization of transport properties in multi-scale systems, 
such as models 
of turbulent fluids, is a delicate task, 
especially when exponents of scaling 
laws and/or universal constants are to be measured from 
Lagrangian statistics. Additional difficulties arise in all cases 
where the standard asymptotic quantities, for example the 
 diffusion coefficients, cannot 
be computed correctly, for limitations due essentially to the finite size of 
the domain and to finite spatio-temporal resolution of the data.    
As we have seen in the previous subsection for the LES trajectories, 
the mean square relative dispersion,  
 seen as a function of time, is generally affected by   
 overlap effects between 
different regimes. 
We therefore use a mathematical tool known as Finite-Scale Lyapunov Exponent, 
 briefly FSLE, a  
technique based on exit-time statistics at fixed scale of trajectory 
separation, formerly introduced in the framework of chaotic dynamical systems 
theory (for a review see Boffetta et al., 2000, and references therein). \\
A dynamical system consists, basically,  
of a $N$-dimensional state vector ${\bm x}$, having 
a set of $N$ observables as components evolving in the so-called phase space, 
and of a $N$-dimensional evolution operator ${\bm F}$, related by a 
first-order  ordinary differential equations system:
\begin{equation} 
{\dot{\bm x} }(t) = {\bm F}[{\bm x}] .
 \label{eq:dynsys}
\end{equation}
If ${\bm F}$ is nonlinear, the system (\ref{eq:dynsys}) can have chaotic 
solutions, that is  limited predictability, 
for which case an infinitesimally small error  
$\delta{\bm x}$ on a trajectory ${\bm x}$ is exponentially amplified in time:
\begin{equation}
\delta{\bm x}(t) \sim \delta{\bm x}(0) \exp{\lambda t}
 \label{eq:lyapexp}
\end{equation} 
 with a (mean) growth rate $\lambda$ 
known as Maximum Lyapunov Exponent (MLE).   
The FSLE is based on the idea of characterizing the growth rate of a  
trajectory perturbation in the whole 
range of scales from infinitesimal to macroscopic sizes. 
In the Lagrangian description of fluid motion, the vector 
 ${\bm x}$ is the tracer trajectory, the operator 
${\bm F}$ is the velocity field, and the error $\delta{\bm x}$ is the distance 
between two trajectories. It is therefore straightforward to consider 
 the relative dispersion of Lagrangian trajectories   
as a problem of finite-error predictability.   

At this regard, 
the FSLE analysis has been applied in a number of recent works as diagnostics 
of transport properties in geophysical systems 
(see, for example, Lacorata et al., 2001; Joseph and Legras, 2002; 
LaCasce and Ohlmann, 2003). 

The procedure to define the FSLE is the following. 
Let $r=|\delta{\bm x}|$ be the distance between two trajectories. 
Given a series of $N$ spatial scales, or thresholds, 
$\delta_1, \delta_2, \cdots ,\delta_N$ have been properly chosen 
such that $\delta_{i+1}=\rho \cdot \delta_i$, for $=1,\cdots ,N-1$  
and with $\rho > 1$, the FSLE is defined as
\begin{equation}
\lambda(\delta) = {{\rm ln} \, \rho  \over  \langle  T(\delta)  \rangle}
\label{eq:fsle}
\end{equation}
where  $\langle T(\delta) \rangle$ is the mean exit-time of $r$ from the 
threshold $\delta=\delta_i$, in other words the mean time taken for $r$ 
to grow from $\delta$ to $\rho \delta$. The FSLE depends very weakly 
on $\rho$ if $\rho$ is chosen not much larger than $1$. The factor $\rho$ 
cannot be arbitrarily close to $1$ because of finite-resolution problems 
and, on the other hand, must be kept sufficiently small in order to avoid 
contamination effects between different scales of motion.   
In our simulations we have fixed $\rho=\sqrt{2}$. 
For infinitesimal $\delta$, the FSLE coincides with the MLE.  
In general, for finite $\delta$,    
the FSLE is expected to follow a power law of the type:  
\begin{equation}
\lambda(\delta) \sim \delta^{-2/\gamma}
\label{eq:powerlaw}
\end{equation}
where the value of $\gamma$ defines the dispersion regime at scale 
$\delta$, for example: 
$\gamma=3$ refers to Richardson diffusion within the 
turbulence inertial range; 
$\gamma=1$ corresponds to standard diffusion, 
that is large-scale uncorrelated spreading of particles. 
These scaling laws can be explained by dimensional argument: 
if the scaling law of the relative dispersion in time is of the form 
$r^2(t) \sim t^{\gamma}$, the inverse of time as function of space gives the 
corresponding scaling (\ref{eq:powerlaw}) of the FSLE.   
In our case, indeed, we seek for a power law related to Richardson diffusion, 
inside the inertial range of the LES: 
\begin{equation}
\lambda(\delta) = \alpha \delta^{-2/3}
\label{eq:powerles}
\end{equation}
where $\alpha$ is a constant depending on the details of the numerical 
experiment.  
The corresponding mean square relative separation is expected 
to follow Eq.~(\ref{eq:2.3}). 
A formula can be derived, which relates the FSLE to the 
Richardson's constant 
(Boffetta and Sokoloff, 2002):
\begin{equation}
C_2 =  \beta {\alpha^3 \over \epsilon} \left ({\rho^{2/3}-1 \over \rho^{2/3} 
{\rm ln} \, \rho}\right )^3 
\label{eq:boffi}
\end{equation}
where $\beta$ is a numerical coefficient equal to $1.75$,  
$\epsilon$ is the energy dissipation measured from the LES and $\alpha$ 
 comes from the best fit of Eq.~(\ref{eq:powerles}) to the data. 
Information about the existence of the inertial range is also given by a 
quantity related to the FSLE, the mean relative Lagrangian velocity at 
fixed scale that we indicate with 
\begin{equation}
\nu(\delta)  =  [\langle \delta {\bm v}(\delta)^2 \rangle ]^{1/2}
\label{eq:velgrad}
\end{equation}
where  
\begin{equation}
 \delta {\bm v}(\delta)^2  = ({\dot{\bm x}^{(1)}} -{\dot{\bm x}^{(2)}})^2  
\label{eq:veldiff}
\end{equation}
is the square  
(Lagrangian) velocity difference between two trajectories, ${\bm x}^{(1)}$ 
and ${\bm x}^{(2)}$,  
on scale $\delta$, that is  for $|{\bm x}^{(1)}-{\bm x}^{(2)}|=\delta$. 
The quantity $\nu(\delta)/ \delta$  is dimensionally equivalent to 
 $\lambda(\delta)$, and, in conditions of sufficient isotropy, it represents 
the spectrum of the relative dispersion rate in real space.   
A scaling law of the type
\begin{equation}
{\nu(\delta) \over \delta } \sim  \delta^{-2/3} 
\label{eq:velgrad2}
\end{equation}
is compatible with the FSLE inside the inertial range and 
therefore with the expected behavior of the turbulent 
velocity difference as function of the scale. 
In Fig.~4(a) we can see, indeed, that
the FSLE measured from the LES data follows the 
behavior of Eq.~(\ref{eq:powerles}), 
from the scale of the spatial resolution 
to about the size of the domain. From the fit we extract the coefficient 
$\alpha=0.1 \; m^{2/3}t^{-1}$. The energy dissipation measured from the LES 
is $\epsilon=6 \cdot 10^{-4} \; m^2 t^{-3}$.  
The formula of Eq.~(\ref{eq:boffi}) gives a measure of the Richardson's 
constant $C_2 \sim 0.5$, affected, at most, by an estimated error of
$\pm 0.2$. 
%$40\%$ 
%relative error.    
In Fig.~4(b) we see, also, that  $\nu(\delta)/\delta$ has been found 
very close to the   
 behavior predicted by Eq.~(\ref{eq:velgrad2}).
\begin{figure}
\begin{center}
\includegraphics[width=12cm,height=8cm]{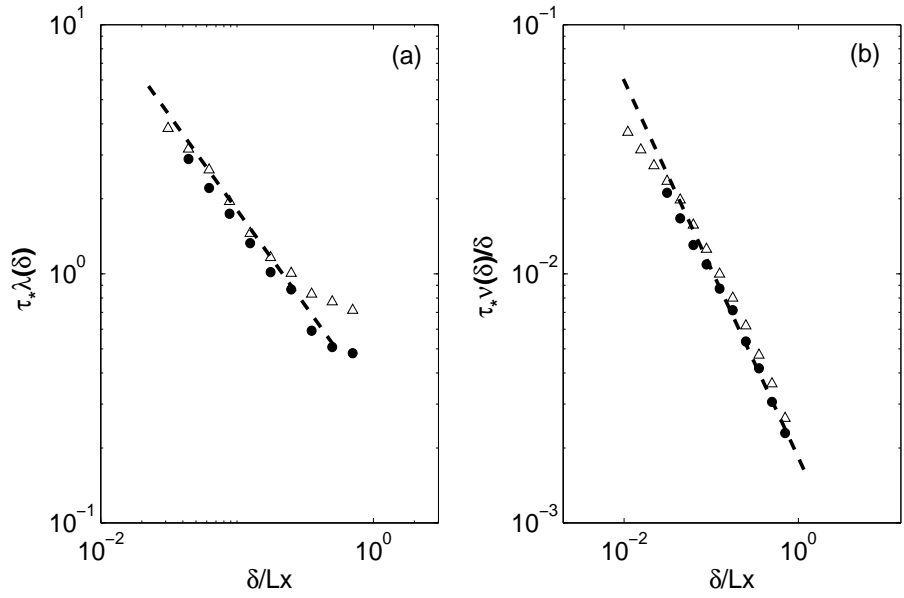}
\caption{a) FSLE at two different resolutions. Triangles: $128^3$
grid points;  Circles: $96^3$ grid points.
The dashed line corresponds to  $\alpha \delta^{-2/3}$ 
with $\alpha=0.1$ $m^2 t^{-3}$. 
b) the same as in a) but for the relative velocity. The dashed 
line has slope $-2/3$.}
\end{center}
\end{figure}
Variations within the error bars are observed by varying the spatial
resolution from $128^3$ grid points (triangles in Fig.~4) to
 $96^3$ grid points (circles).

\section{Conclusions and perspectives}
We have investigated the problem of relative dispersion in a
neutrally stratified  planetary boundary layer simulated by means of
Large-Eddy Simulations. In particular, our attention has been 
focused on the possible emergence of the celebrated Richardson's
law ruling the separation in time of particle pairs.\\ 
The difficulties in observing such behavior in a realistic PBL
mainly rely on the fact that it is hard to obtain a PBL with 
 a sufficiently extended inertial range of scales. 
For this reason, standard techniques 
to isolate the Richardson's law and the relative
constant turn out to be inconclusive, the results being 
strongly dependent, for instance, on the choice of the initial
pair separations. To overcome this problem,
we have applied, for the first time in the context of boundary layer physics,
a recently established technique coming from the study of 
dynamical systems.  As a result, a clean region of scaling showing
the occurrence of the Richardson law has been observed and an
accurate, systematic,  measure of the Richardson constant became possible.
Its value is $C_2=(0.5\pm 0.2)$, where the error bar has been determined
in a very conservative way. Such estimation is compatible with
the one obtained from Fig.~3 in the case of 
initial pair separation equal to $\Delta x$.
The important point is that the new strategy gives a result that, 
by construction, does not depend on the initial pair separations.
As already emphasized this is not the case for the standard
approach.

Clearly, our study is not the end of the story. The following
points appear to be worth investigating in a next future.\\
The first point is related to the fact that in our simulations
we did not use any sub-grid model for the unresolved Lagrangian
motions. The main expected advantage of  SGS Lagrangian 
parameterizations is to allow  the choice of initial pair 
 separations smaller than 
the grid mesh spacing, a fact that would cause a reduction of the 
crossover from initial conditions to the genuine $t^3$ law.
The investigation of this important point is left for future research.\\
Another point is related to the investigation of the 
probability density function (pdf) 
of pair separation. In the present study, we have focused 
on the sole second moment of this pdf. There are, indeed,
several solutions for the diffusion equation (\ref{diffusion})
all giving pdfs compatible with the $t^3$ law. 
The solution for the pdf essentially depends on the choice
for the eddy-diffusivity field, $D(r)$.  The answer to this question
 concerns applicative studies related, for example, 
to pollutant dispersion because of the importance of correctly
describing the occurrence of extreme, potentially dangerous, 
events.\\
Finally, it is also interesting to investigate whether or not
the Richardson law rules the behavior of pair separations
 also in buoyancy-dominated  
boundary layers. In this case, the role
of buoyancy could modify the expression for the  eddy-diffusivity field, 
$D(r)$, thus giving rise to an essentially new regime which is however
up to now totally unexplored.

\section*{Aacknowledgements}
This work has been partially supported by Cofin 2001, prot. 2001023848
(A.M.) and by CNPq 202585/02   (E.P.M.F.).
We acknowledge useful discussions with Guido Boffetta and Brian Sawford.


\begin{thebibliography}{}

\bibitem{AMR03}
Antonelli, M., A.~Mazzino and U.~Rizza.
\newblock {Statistics of temperature fluctuations 
in a buoyancy dominated boundary layer flow simulated by a 
Large-eddy simulation model}.
\newblock {\em J. Atmos. Sci.}, 60:215--224, 2003.


\bibitem{BC00}
Boffetta, G., A. Celani.
\newblock { Pair dispersion in turbulence}.
\newblock {\em Physica A}, 280:1--9, 2000.

\bibitem{BS02}
Boffetta G. and I.M. Sokolov. 
\newblock {Relative dispersion in fully developed turbulence:
the Richardson's law and intermittency corrections}.
\newblock {\em Phys. Rev. Lett}, 88:094501, 2002

\bibitem{BCCLV00}
Boffetta, G., A. Celani, M. Cencini, G. Lacorata and A. Vulpiani.
\newblock {Non Asymptotic Properties of Transport and Mixing}.
\newblock {\em Chaos}, 10:1--9, 2000.

\bibitem{EM96}
Elliot F.W. and A.J. Majda.
\newblock {Pair dispersion over an inertial range
spanning many decades}. 
\newblock {\em Phys. Fluids}, 8:1052--1060, 1996.


\bibitem{F95}
Frisch, U.
\newblock {Turbulence: the legacy of A.N. Kolmogorov}.
\newblock Cambridge University Press, 1995.

\bibitem{FV98}
Fung J.C.H. and J.C~Vassilicos.
\newblock {Two-particle dispersion in turbulent-like flows}. 
\newblock {\em Phys. Rev. E}, 57:1677--1690, 1998.

\bibitem{JL02}
Joseph B. and B.~Legras. 
\newblock {Relation between Kinematic Boundaries, 
Stirring and Barriers for the Antarctic Polar Vortex}. 
\newblock {\em J. Atmos. Sci}, 59:1198--1212, 2002.

\bibitem{LO03}
LaCasce J.H. and C.~Ohlmann. 
\newblock {Relative Dispersion at the Surface of the 
Gulf of Mexico}. 
\newblock {\em J. of Mar. Res.}, submitted, 2003.

\bibitem{LAV01}
Lacorata, G., E. Aurell and A. Vulpiani. 
\newblock {Drifter Dispersion in the 
Adriatic Sea: Lagrangian Data and Chaotic Model}. 
\newblock {\em Ann. Geophys.}, 19:121--129, 2001.

\bibitem{M84} 
Moeng, C.-H.
\newblock {A large-eddy-simulation model for the study of 
planetary boundary-layer turbulence}.
\newblock {\em  J. Atmos. Sci.}, 41:2052--2062, 1984.

\bibitem{MS94}  
Moeng C.-H., and P.P.~Sullivan.
\newblock {A comparison of shear and 
buoyancy driven Planetary Boundary Layer flows}.
\newblock {\em J. Atmos. Sci.}, 51:999--1021, 1994.

\bibitem{MJ75}
Monin, A.S. and Yaglom A.M. 
\newblock {Statistical Fluid Mechanics: 
Mechanics of Turbulence}. 
\newblock Cambridge, MA/London, UK: MIT, 1975.

\bibitem{OM00}
Ott, S. and J. Mann.
\newblock {An experimental investigation of the relative diffusion
of particle pairs in three-dimensional turbulent flow}.
\newblock {\em J. Fluid Mech.}, 422,:207--223, 2000.

\bibitem{R26}
Richardson, L.F.
\newblock {Atmospheric diffusion shown on a distance-neighbor 
graph}.
\newblock {\em Proc. R. Soc. London Ser. A},  110:709--737, 1926.

\bibitem{S01}
Sawford B.
\newblock {Turbulent relative dispersion}.
\newblock {\em Ann. Rev. Fluid Mech.}, 33:289--317, 2001.

\bibitem{SMM94} 
Sullivan, P.P., J.C.~McWilliams, and C.-H.~Moeng. 
\newblock {A sub-grid-scale model for large-eddy simulation of 
planetary boundary layer flows}. 
\newblock {\em Bound. Layer Meteorol.}, 71:247--276, 1994.

\end{thebibliography}
\end{document}